\documentclass[showpacs,amsmath,amssymb,aps,prd,groupedaddress,superscriptaddress,nofootinbib]{revtex4-2} 
\usepackage[utf8]{inputenc}
\usepackage{graphicx}
\usepackage{tensor}
\usepackage{bm}
\usepackage{color}
\usepackage{array}
\usepackage{tabularx}

\newcommand{\rcd}[1]{\tensor{\mathring{\nabla}}{#1}}
\newcommand{\leviconnection}[1]{\tensor{\mathring{\bm{\nabla}}}{#1}}
\newcommand{\rsconnection}[1]{\tensor{\mathring{\omega}}{#1}}

\newcommand{\potential}[1]{\tensor{\Sigma}{#1}}
\newcommand{\calt}{{\cal T}}
\newcommand{\pd}[1]{\tensor{\partial}{#1}}
\newcommand{\nablab}{\bm{\nabla}\!}

\newcommand{\e}[2][]{\tensor{#1{e}}{#2}}

\newcommand{\teta}[2][]{\tensor{#1\vartheta}{#2}}

\newcommand{\torsion}[1]{\tensor{T}{#1}}

\newcommand{\energy}[1]{\tensor{t}{#1}}

\newcommand{\acceleration}[2][]{\tensor{#1\phi}{#2}}

\begin{document}


\title{Gravitational energy in $pp$-wave spacetimes}



\author{Costa, R. D.}

\author{J. B. Formiga}
\email[]{jansen@fisica.ufpb.br}
\affiliation{Departamento de Física, Universidade Federal da Paraíba, Caixa Postal 5008, 58051-970 João Pessoa, Pb, Brazil}


\begin{abstract}
The description of the gravitation energy is a long standing problem. Although some success has been achieved, there is no satisfactory solution to this problem yet. Probably the most promising approach to this problem is given by teleparallelism. Many consistent and interesting results have been obtained in the context of the so-called Teleparallel Equivalent of General Relativity, including results obtained with spacetimes that are neither static nor asymptotically flat. One example is the analysis of the plus-polarized $pp$-waves that has been made recently [Phys. Rev. D 108, 044043 (2023)]. In this paper, this analysis is extended to arbitrary polarization and some new results that shows the consistency of the teleparallel approach is obtained.
\end{abstract}


\maketitle

\section{Introduction} 
The notions of energy and momentum are among the most important concepts of physics. However, they are not always well defined in general relativity (GR), and many attempts have been made to circumvent this problem. A possible solution to this problem is given by the so-called Teleparallel Equivalent of General Relativity (TEGR), which allows for a improved treatment of the energy problem, both globally and locally. In this paper, we study the $pp$-wave solutions with both plus and cross polarizations, generalizing the results obtained in Ref.~\cite{PhysRevD.108.044043}, and obtain some new results that show the consistency of the TEGR.

Energy conservation,  energy flow and the like are used to study the properties of many different systems. The notion of energy is fundamental to the evolution of a system; in quantum mechanics, for example, the Hamiltonian is essential to  the Schr{\"o}dinger equation, which governs the time evolution of the state vector. A well-defined energy is also required for statistical thermodynamics \cite{CRovelli1993}, and useful for gravitation theory and relativistic astrophysics \cite{Gravitation}. The list could go on and on but the point would be the same, the concept of energy is too important for physics and we should not give up on a well-defined gravitation energy so easily.

There have been many proposals for the gravitational energy-momentum tensor in GR \cite{TheBerlinYearsVolume6p98,10.2307/20488488,PhysRev.89.400,Weinberg:1972kfs,Landaufourthv2,OZKURT2017}, but all of them depend on the coordinate system  and, because of that, they are usually called pseudotensors. 
One of the problems concerns the localization of the gravitational energy, with the principle of equivalence and the notion of locality playing an important role in the problem  (see, e.g., Sec.~10 of Ref.~\cite{NORTON1985203} and Sec.~V of Ref.~\cite{PhysRevD.106.044021}; see also Refs.~\cite{doi:10.1119/1.10744} and \cite{doi:10.1119/1.4895342}). The strongest arguments against a localized gravitational energy can be found in section~20.4 of Ref.~\cite{Gravitation}. They say that there is no unique formula for the localized energy, this energy has no weight and it is not observable; Einstein's equivalence principle forbids the localization of the gravitational energy. Although GR admits a global definition of energy and momentum in asymptotically flat spacetimes, the so-called ADM energy, this theory fails to provide a coordinate independent definition of this energy; it also fails to provide a unique definition of a quasi-local energy \cite{Penrose1982,PhysRevD.47.1407}. Furthermore, the restriction to asymptotically flat spacetimes is too strong; it excludes, for example, cosmological solutions. 

On the other hand, there is still hope that the above arguments against a well-defined energy can be circumvent in a different approach to gravity, preferably one that does not deviate too much from GR. The best candidate seems to be the TEGR, since its energy-momentum tensor has many advantages over the pseudotensors of GR \cite{ANDP:ANDP201200272,PhysRevD.106.044021}: this tensor is coordinate independent, compatible with a massless field (traceless), and allows for a $4$-momentum that generalizes the ADM one. The TEGR energy-momentum tensor also opens the door to the possibility of solving the apparent conflict between a gravitational energy density and the principle of equivalence (see, e.g., Sec.~III. D. of Ref.~\cite{PhysRevD.108.044043} and Sec.~3.2 of Ref.~\cite{doi:10.1142/S0217732322502224}). In addition, TEGR seems to have a better action principle than GR: it does not require additional surface terms and admits the path-integral approach to quantization \cite{PhysRevD.96.044042}.

In Ref.~\cite{PhysRevD.108.044043} the gravitational energy problem was analyzed from many different perspectives, including both local and global viewpoints. However, the results were limited to plus-polarized gravitational waves; the electromagnetic wave was also restricted to fixed polarization. Here, we remove those constraints and generalize the results in Ref.~\cite{PhysRevD.108.044043}. In addition, we obtain the relations of the spacetime $4$-momentum and the gravitational energy density with the expansion tensor associated with the timelike geodesic congruence used to build the frame.

The next section is devoted to a brief review of the TEGR, and also to write down Einstein-Maxwell equations in the teleparallel form. Our main results are in Sec.~\ref{04122024b}, while some final remarks are in Sec.~\ref{04122024c}. The relation between the coordinates used here and a coordinate system that has a closer relationship with the amplitudes of the waves is given in the appendix.

\section{Teleparallelism}\label{04122024a}
The notion of teleparallelism is rather intuitive, it is the idea that the parallel transport of a vector is path independent, and is present in any spacetime whose curvature tensor vanishes \cite{Eisenhart1927}. Minkowski spacetime is a trivial example of such a spacetime; it has nice properties such as a global inertial frame of reference, the Poincar\'e symmetry group, well-defined conservation laws etc. However, the geometry in which GR is formulated does not have teleparallelism, and many of those desirable properties are lost. One way to tackle this problem is to formulate a geometry with at least two affine connections, one to deal with the non-triviality of the metric (Levi-Civita connection) and other to deal with the distant parallelism (Weitzenb\"{o}ck connection).

Let us denote the tetrad field by $\{\e{_a}\}$ (frame) and $\{\teta{^a}\}$ (coframe), where Latin indices from the beginning of the alphabet run from $(0)$ to $(3)$ and represent flat Minkowski indices [$\e{_a}\cdot\e{_b}\equiv g(\e{_a},\e{_b})=\eta_{ab}=\textrm{diag}(-1,+1,+1,+1)$]; we use Greek letters to represent the coordinate indices, running over  $0,1,2,3$, and Latin indices in the middle of the alphabet to represent the spatial part, running over $1,2,3$. Using these conventions, we can write the coefficients of the Levi-Civita connection $\leviconnection{}$ in the tetrad basis\footnote{In the coordinate basis, the connection coefficients are the Christoffel symbols.} as
\begin{align}
\rsconnection{^a_b_c}=\frac{1}{2}\left( \torsion{_b_c^a}+\torsion{_c_b^a}-\torsion{^a_b_c} \right), \label{10092020a}
\end{align}  
where
\begin{align}
\torsion{^a_\mu_\nu}=\pd{_\mu}\e{^a_\nu}-\pd{_\nu}\e{^a_\mu} \label{06122021a}
\end{align}
is the object of anholonomity and  $\e{^a_\nu}$ are the components of the tetrad field in the coordinate basis $\{dx^\nu\}$. [The components of the frame is denoted by $\e{_a^\mu}$ and satisfy the equations $\e{^a_\mu}\e{_b^\mu}=\delta^a_b$ and $\e{^a_\nu}\e{_a^\mu}=\delta^\mu_\nu$.] The impossibility of finding a frame in which $\rsconnection{^a_b_c}=0$ everywhere is a consequence of the non-triviality of the metric tensor (the absence of a coordinate system in which $g_{\mu\nu}$ is constant  everywhere), and implies that the curvature tensor of the Levi-Civita connection does not vanish. Therefore, there is no distant parallelism with that affine connection, and the ordinary description of GR cannot be teleparallel.

Nonetheless, we can add to the GR framework a connection $\nablab$ such that, by definition, we have $\nablab{_\mu}\e{_a}=0$ everywhere\footnote{Being a parallelizable manifold is considered to be a strong global topological restriction, but it is a highly desirable property.}. This new connection must have either torsion or nonmetricity (or both). Here, we choose to work with a connection that is metric compatible (vanishing nonmetricity), in which case the torsion is responsible for ensuring the validity of $\nablab{_\mu}\e{_a}=0$. (We call the frame in which this equation holds {\it the teleparallel frame}.) This connection is known as the Weitzenb\"{o}ck connection, and its torsion is called Weitzenb\"{o}ck torsion. In the teleparallel frame, the Weitzenb\"{o}ck torsion coincides with the object of anholonomity and is given by Eq.~\eqref{06122021a}. [As it is clear in Eq.~\eqref{10092020a}, the Levi-Civita connection is uniquely determined by this torsion and vice versa.]

We also restrict ourselves to the TEGR \cite{ANDP:ANDP201200272}, whose field equations are exactly Einstein's equations. At the classical level, it differs from GR only at the level of the action and of the geometrical setup. This, however, is enough to yield a better road to the solution of the energy problem; it allows, for instance, a generalization of the ADM energy momentum \cite{PhysRevD.106.044021}.

Although some variation of the definition of energy can be found in the literature of teleparallelism [compare\footnote{Some authors define the $4$-momentum with a spacetime index. However, as explained in Sec.~VI of Ref.~\cite{PhysRevD.106.044021}, there is no ambiguity because the fundamental variable in the TEGR is $\e{^a_\mu}$, not $g_{\mu\nu}$. Therefore, the momentum canonically conjugated is of the type $\Pi^{a\nu}$, which leads to a $P^a$ rather than to either $P^\mu$ or $P_\mu$.}, for example, Eq.~(3.21) in Ref.~\cite{NesterPositive1989} and Eq.~(35) of Ref.~\cite{ANDP:ANDP201200272}], when one takes the tetrad field as the fundamental variable representing the gravitational field and adopts either the Lagrangian or the Hamiltonian formalism, one finds a unique expression for the spacetime and gravitational $4$-momenta in the teleparallel frame. In turn, this also gives us a unique way to write Einstein's equation by putting the energy-momentum tensor of the gravitational field on the same footing of that of the matter field. The TEGR form of Einstein's equation is
\begin{equation}
\pd{_\alpha}\left( e\potential{^a^\mu^\alpha}\right)=\frac{e}{4k}\left(\energy{^\mu^a}+{\cal T}^{\mu a}\right), \label{29032019k}
\end{equation}
where $k=c^4/(16\pi G)$ and $\potential{^a^\mu^\alpha}$ is called {\it superpotential}, which can be given by \cite{doi:10.1142/S0219887824502116,PhysRevD.73.124017}
\begin{align}
\potential{_a_b_c}=\frac{1}{2}\rsconnection{_c_a_b}+\rsconnection{^d_d_{[c}}\tensor{\eta}{_{b]a}}.
\label{14082024h}
\end{align} 
${\cal T}^{\mu a}$ is the matter energy-momentum tensor, while $\energy{^\mu^a}$ is interpreted as the gravitational energy-momentum tensor. According to this view,  we interpret $\tau^{\mu a}=\energy{^\mu^a}+{\cal T}^{\mu a}$ as the spacetime energy-momentum tensor, which is supposed to represent the content of energy and momentum locally, thus realizing a kind of localization  of the energy. In terms of the Levi-Civita connection coefficients, $\energy{^\mu^a}$ takes the form
\begin{align}
\energy{^b_a}= 2k\Bigl(2\rsconnection{^c_{[ad]}}\rsconnection{^b_c^d}-2\rsconnection{^b_{[ad]}}\rsconnection{^c_c^d}
-\rsconnection{^c_c_a}\rsconnection{^d_d^b}
\nonumber\\
+\delta^{b}_{a}\rsconnection{^c_{[c|f}}\rsconnection{^d_{|d]}^f}  \Bigr).
\label{09092024d}
\end{align} 
(For the original expressions, i.e, those written in terms of the torsion components, see Maluf \cite{ANDP:ANDP201200272}.) The tensor $\energy{^a^b}$ does not depend on the coordinate system and its dependency of the teleparallel basis $\{\e{_a}\}$ is analogous to that of the $4-$acceleration (both depend on the congruence of curves because they reflect its properties).

In GR, the energy conservation is written as $\rcd{_\mu}{\cal T}^{\mu\nu}=0$, and the tensor ${\cal T}^{\mu\nu}$ describes the local energy content of nongravitational fields. There are at least two ways to see that ${\cal T}^{\mu\nu}$ does not account for all the energy content. The first one comes from physical systems, such as gravitational waves, which carry energy, and some thought experiments, such as those of Feynman \cite{universe2030022}, which leaves no doubt about the existence of the gravitational energy. The second way,  as pointed out by Penrose \cite{Penrose1982}, is that the presence of the gravitation energy in $\rcd{_\mu}{\cal T}^{\mu\nu}=0$ is manifested in the fact that this conservation law does not give rise to an integral conservation law. In the context of the TEGR, however, a meaningful conservation law can be obtained from Eq.~\eqref{29032019k} by simply applying $\partial_\mu$ on both sides, which yields $\partial_\mu(e\tau^{\mu a})=0$, since $\potential{^a^\mu^\nu}=-\potential{^a^\nu^\mu}$. A natural integral conservation law emerges from Eq.~\eqref{29032019k} \cite{ANDP:ANDP201200272}.

Another important quantity that can be obtained from Eq.~\eqref{29032019k} is the quasi-local $4$-momentum $P^a$ over a two-surface. In the TEGR, the $4$-momenta of the matter field $P^a_M$, gravitational field $P^a_g$, and the spacetime $P^a$ are naturally defined as 
\begin{align}
P^a_M\equiv\int_V d^3x e{\cal T}^{0a},\quad P^a_g\equiv\int_V d^3x e\energy{^0^a},\quad P^a\equiv\int_V d^3x e\tau^{0a}.
\label{02012023b}
\end{align}
where the integrals are over a region $V$, a spacelike three-surface defined by $x^0=$ constant. (Note that $P^a= P^a_g+P^a_M$.) Now, if the region $V$ is free from singularities, then, by using Eq.~\eqref{29032019k}, one finds that
 \begin{align}
P^a=4k\oint_S dS_i e\Sigma^{a0i}.
\label{29102023c}
\end{align}
The integral above is over the boundary of $V$. In some cases, this boundary will be the combination of many two-surfaces that exclude the singularities from the region $V$. If the spacetime is asymptotically flat, the total spacetime energy is taken as the integral over the external boundary. (For more details about the latter point, see section II C. of Ref.~\cite{PhysRevD.108.044043}.)

There is a rich literature on teleparallelism and the interested reader is referred to \cite{SAUER2006399,MOLLER1961118,PhysRevD.14.2521,PhysRevD.19.3524,ANDP:ANDP201200272} for further details. See also Refs.~\cite{NesterPositive1989,doi:10.1063/1.530774,Maluf1999,PhysRevD.64.084014,PhysRevD.65.124001,PhysRevD.82.124035,Oko_w_2013,Oko_w_2014,PhysRevD.94.104045,Blixt2021,NesterSTGR1999,PhysRevD.98.044048}. (The reader familiar with the Metric-Affine Gravity may be interested in Sec.~5.9 of Ref.~\cite{HEHL19951} and Sec.~3.3 of Ref.~\cite{blagojevic2002gravitation}.)

\subsection{Einstein-Maxwell equations in the teleparallel form}
There is not much difference in Einstein-Maxwell equations in the TEGR: the only difference is in the form in which Einstein's equation is written, Eq.~\eqref{29032019k}. Maxwell's equations in TEGR can be written in the familiar form of GR:
\begin{align}
\rcd{_\nu} F^{\mu\nu}=c\mu_0J^\mu,\ \partial_{[\alpha}F_{\beta\gamma]}=0,
\label{24092024a}\\
F_{\mu\nu}=\partial_\mu A_\nu-\partial_\nu A_\mu, 
\label{24092024c}
\end{align}
where $\rcd{_\nu}$ is the Levi-Civita covariant derivative, and $F_{\mu\nu}$ is the electromagnetic tensor. In turn, Maxwell energy-momentum tensor can be written as
\begin{align}
\tensor{\calt}{^\mu^\nu}=\epsilon_0(\tensor{F}{^\mu^\alpha}\tensor{F}{^\nu_\alpha}-\frac{1}{4}g^{\mu\nu}F_{\alpha\beta}F^{\alpha\beta}).
\label{24092024b}
\end{align}
[We are using SI units. Note that we have chosen to work with $F_{\mu\nu}$ having units of electric field, which means that $F_{01}=-E_1$ and $F_{12}=cB_3$. Equation \eqref{24092024b} with $F_{\mu\nu}$ in units of magnetic field is obtained by exchanging $\epsilon_0$ for $1/\mu_0$, in which case $F_{01}=-E_1/c$ and $F_{12}=B_3$.] Equations \eqref{29032019k}, \eqref{24092024a}, and \eqref{24092024b} are the Einstein-Maxwell equations.

\section{Plane waves}\label{04122024b} 

The line element for the so-called $pp$-waves can be written as
\begin{align}
ds^2=-c^2dt^2+l_x^2(u)dx^2+l_y^2(u)dy^2+2l_{xy}dxdy+dz^2,
\label{16022024d}\\
l_x^2=f_1^2(u)+g_2^2(u),\ l_y^2=f_2^2(u)+g_1^2(u),
\nonumber\\
l_{xy}=f_1(u)f_2(u)+g_1(u)g_2(u),\ u=t-z/c,
\label{16022024e}
\end{align}
where the functions $f_1$, $f_2$, $g_1$, and $g_2$ must satisfy Eq.~\eqref{26082024e} \cite{GriffithsJBWaves}. (For another coordinate system and the differences between the conventions used here and those of Ref.~\cite{GriffithsJBWaves}, see the Appendix.)

\subsection{Teleparallel frame and the associated energy}
In order to make the calculations simpler and easier to handle, let us proceed as follows. One can always find an orthonormal basis $\{\hat{t},\hat{x},\hat{y},\hat{z}\}$ such that the coframe and frame can be given in the form:
\begin{align}
\e{^a_\mu}=-\hat{t}^a\hat{t}_\mu+\hat{x}^a\hat{x}_\mu+\hat{y}^a\hat{y}_\mu+\hat{z}^a\hat{z}_\mu,
\label{13082024a}\\
\e{_a^\mu}=-\hat{t}_a\hat{t}^\mu+\hat{x}_a\hat{x}^\mu+\hat{y}_a\hat{y}^\mu+\hat{z}_a\hat{z}^\mu,
\label{13082024b}
\end{align}
in which $(\hat{t}^\mu,\hat{x}^\mu,\hat{y}^\mu,\hat{z}^\mu)$ are the components in the coordinate basis $\{\partial_\mu\}$, and $(\hat{t}^a,\hat{x}^a,\hat{y}^a,\hat{z}^a)$ are the components in a certain frame $\e{_a}$ that satisfies the scalar product  $\e{_a}\cdot\e{_b}=\eta_{ab}=\textrm{diag}(-1,+1,+1,+1)$.

The metric components $g_{\mu\nu}$ take the simple form
\begin{align}
g_{\mu\nu}=-\hat{t}_\mu\hat{t}_\nu+\hat{x}_\mu\hat{x}_\nu+\hat{y}_\mu\hat{y}_\nu+\hat{z}_a\hat{z}_\nu,
\label{13082024c}
\end{align}
which is analogous to the metric tensor written in the null tetrads of the Newman-Penrose formalism [see e.g. Eq.~(2.3a) in Ref.~\cite{10.1063/1.1724257}]. Note, however, that here $\hat{t}$ is timelike and the others are spacelike: 
\begin{align}
\hat{t}\cdot\hat{t}=-1,\ \hat{x}\cdot\hat{x}=\hat{y}\cdot\hat{y}=\hat{z}\cdot\hat{z}=1,
\label{13082024d}\\
\hat{t}\cdot\hat{x}=\hat{t}\cdot\hat{y}=\hat{t}\cdot\hat{z}=\hat{x}\cdot\hat{y}=\ldots=0.
\label{13082024e}
\end{align}

We have not yet fixed the frame $\e{_a}$ and the components of $\{\hat{t},\hat{x},\hat{y},\hat{z}\}$. We do that by choosing a tetrad that is adapted to the coordinate system $(ct,x,y,z)$. A possible choice\footnote{It is clear that any frame adapted to the coordinates $(ct,x,y,z)$ must necessarily have $\e{_{(0)}}=\partial_0$ and $\e{_{(3)}}=\partial_3$. However, since it is not possible to take $\e{_{(1)}}\propto\partial_1$ and $\e{_{(2)}}\propto\partial_2$, due to the orthonormality condition, there is an ambiguity in the choice of these vector fields.} is given by Eqs.~\eqref{13082024f}, \eqref{13082024h} and \eqref{13082024i} below:
\begin{align}
\hat{t}^a=\delta^a_{(0)},\ \hat{x}^a=\delta^a_{(1)},\ \hat{y}^a=\delta^a_{(2)},\ \hat{z}^a=\delta^a_{(3)},
\label{13082024f}\\
\hat{t}_a=-\delta_a^{(0)},\ \hat{x}_a=\delta_a^{(1)},\ \hat{y}_a=\delta_a^{(2)},\ \hat{z}_a=\delta_a^{(3)},
\label{13082024g}
\end{align}
\begin{align}
\hat{t}^\mu=\delta^\mu_0,\ \hat{x}^\mu=\frac{g_1}{e}\delta^\mu_1-\frac{g_2}{e}\delta_2^\mu, 
\label{13082024h}\\
\hat{y}^\mu=-\frac{f_2}{e}\delta^\mu_1+\frac{f_1}{e}\delta_2^\mu,\ \hat{z}^\mu=\delta^\mu_3,
\label{13082024i}\\
\hat{t}_\mu=-\delta_\mu^0,\ \hat{x}_\mu=f_1(u)\delta_\mu^1+f_2(u)\delta^2_\mu,
\label{13082024j}\\
\hat{y}_\mu=g_2(u)\delta_\mu^1+g_1(u)\delta^2_\mu,\ \hat{z}_\mu=\delta_\mu^3,
\label{13082024k}
\end{align}
where
\begin{equation}
e=f_1g_1-f_2g_2
\label{16022024f}
\end{equation}
is the determinant of the tetrad field $\e{^a_\mu}$. Notice that the choice of $\e{_a}$ is implicit in the relations $\hat{t}^\mu=\e{_a^\mu}\hat{t}^a$, $\hat{x}^\mu=\e{_a^\mu}\hat{x}^a$ etc. 

Notice also that, by using Eq.~\eqref{13082024f}, we have chosen $\{\hat{t},\hat{x},\hat{y},\hat{z}\}$ in such a way that they coincide with $\{\e{_{(0)}},\e{_{(1)}},\e{_{(2)}},\e{_{(3)}}\}$ [for example, $\hat{x}=\hat{x}^a\e{_a}=\delta^a_{(1)}\e{_a}=\e{_{(1)}}$]. This kind of choice does not have to be always the case, because there are situations in which the most convenient choice will give different bases. In any case, it is still useful to use the notation $\{\hat{t}^\mu,\hat{x}^\mu,\hat{y}^\mu,\hat{z}^\mu\}$ in place of $\{\e{_{(0)}^\mu},\e{_{(1)}^\mu},\e{_{(2)}^\mu},\e{_{(3)}^\mu}\}$.

We can take advantage of the symmetries of the problem by writing the torsion components in terms of the components of the vector fields $\hat{t}$, $\hat{x},\hat{y}$, $\hat{z}$. In fact, it also turns out to be useful to write it in terms of the null vector field $v^a\equiv \hat{t}^a+\hat{z}^a$. From Eqs. \eqref{13082024d} and \eqref{13082024e}, we see that 
\begin{align}
v\cdot v=0,\ v\cdot\hat{x}=v\cdot\hat{y}=0,\ v\cdot\hat{t}=-1,\ v\cdot\hat{z}=1.
\label{14082024c}
\end{align}
In turn, from Eqs.~\eqref{13082024j} and \eqref{13082024k}, we obtain the identities
\begin{align}
 \delta_\mu^1=\frac{1}{e}(g_1\hat{x}_\mu-f_2\hat{y}_\mu),\ \delta_\mu^2=\frac{1}{e}(-g_2\hat{x}_\mu+f_1\hat{y}_\mu),
\label{14082024a}\\
\delta_\mu^0=-\hat{t}_\mu,\ \delta_\mu^3=\hat{z}_\mu,\ \partial_\mu u=-\frac{1}{c}v_\mu.
\label{14082024b}
\end{align}
We can use Eqs.~\eqref{13082024j}, \eqref{13082024k}, \eqref{14082024a}, and \eqref{14082024b} in order to write the derivatives of $\hat{x}_\mu$ and $\hat{y}_\mu$ in the convenient form
\begin{align}
\partial_{[\mu}\hat{x}_{\nu]}=\alpha v_{[\mu}\hat{x}_{\nu]}+\bar{\beta}v_{[\mu}\hat{y}_{\nu]},\ \partial_{[\mu}\hat{y}_{\nu]}=\beta v_{[\mu}\hat{x}_{\nu]}+\bar{\alpha}v_{[\mu}\hat{y}_{\nu]},
\label{24092024d}\\
\alpha\equiv \frac{-\dot{f}_1g_1+\dot{f}_2g_2}{ec},\ \bar{\alpha}\equiv \frac{-\dot{g}_1f_1+\dot{g}_2f_2}{ec},
\label{14082024e}\\
\beta\equiv \frac{\dot{g}_1g_2-\dot{g}_2g_1}{ec},\ \bar{\beta}\equiv \frac{\dot{f}_1f_2-\dot{f}_2f_1}{ec}.
\label{14082024f}
\end{align}
(The dot denotes the derivative with respect to $u$.)  From Eq.~\eqref{26082024e}, one can see that  $\beta=\bar{\beta}$; hence, from now on, we use only $\beta$.

We are now in a position to calculate all the teleparallel quantities easily. Let us start with the Weitzenb\"{o}ck torsion. Substituting Eq.~\eqref{13082024a} into \eqref{06122021a} and using Eq.~\eqref{24092024d} (note that $\hat{t}^a$, $\hat{t}_\mu$ , $\hat{z}^a$, $\hat{z}_\mu$ are constant), we find
\begin{align}
\torsion{_a_b_c}=2(\alpha\hat{x}_a+\beta\hat{y}_a)v_{[b}\hat{x}_{c]}+2(\beta\hat{x}_a+\bar{\alpha}\hat{y}_a)v_{[b}\hat{y}_{c]},
\label{14082024d}
\end{align}
 Using Eq.~\eqref{14082024d} in Eq.~\eqref{10092020a}, we find
\begin{align}
\rsconnection{_a_b_c}=-2\left(\alpha\hat{x}_b+\beta\hat{y}_b\right)v_{[a}\hat{x}_{c]}-2\left(\beta\hat{x}_b+\bar{\alpha}\hat{y}_b\right)v_{[a}\hat{y}_{c]}.
\label{14082024g}
\end{align}

The acceleration tensor $\acceleration{_a_b}$ is related to $\rsconnection{_a_b_c}$ by $\acceleration{_a^b}=c\rsconnection{^b_{(0)}_a}$. Using this relation and Eq.~\eqref{13082024g}, we see that Eq.~\eqref{14082024g} yields $\acceleration{_a_b}=0$. Hence, the teleparallel frame  is a freely falling frame with a triad that does not rotate with respect to a Fermi-Walker transported frame.

In obtaining the superpotential from Eq.~\eqref{14082024h}, we calculate $\rsconnection{^d_d_c}$ and use the Minkowski metric written in the form $\eta_{ab}=-\hat{t}_a\hat{t}_b+\hat{x}_a\hat{x}_b+\hat{y}_a\hat{y}_b+\hat{z}_a\hat{z}_b$. Equation \eqref{14082024g} gives
\begin{align}
\rsconnection{^d_d_c}=(\alpha+\bar{\alpha})v_c.
\label{14082024i}
\end{align}
[Notice that $\hat{x}_d\hat{t}^{[d}\hat{x}_{c]}=-(1/2)\hat{t}_c$, $\hat{x}_d\hat{x}^{[d}\hat{z}_{c]}=(1/2)\hat{z}_c$ and so on; where $A^{[a}B_{b]}\equiv (1/2)(A^aB_b-A_bB^a)$.] Substituting Eqs.~\eqref{14082024g} and \eqref{14082024i} into Eq.~\eqref{14082024h}, we find 
\begin{align}
\potential{^a^b^c}=&(-\bar{\alpha}v^{[b}\hat{x}^{c]}+\beta v^{[b}\hat{y}^{c]})\hat{x}^a
\nonumber\\
&+(\beta v^{[b}\hat{x}^{c]}-\alpha v^{[b}\hat{y}^{c]})\hat{y}^a
-(\alpha+\bar{\alpha})\hat{t}^{[b}\hat{z}^{c]}v^a.
\label{05092024a}
\end{align}
From this equation and Eqs.~\eqref{13082024f}, \eqref{13082024h} and \eqref{13082024i}, one can easily find an explicit expression for $\potential{^a^\mu^\nu}$ and, therefore, $\potential{^a^0^i}$.

Let us evaluate Eq.~\eqref{29102023c} over the faces of the rectangular parallelepiped $x_1<x<x_2$, $y_1<y<y_2$, and $z_1<z<z_2$. Since $e\potential{^a^0^i}$ does not depend on either $x$ or $y$, the only contribution is from $\potential{^a^0^3}=-(1/2)(\alpha+\bar{\alpha})v^a$ evaluated over the faces $z=z_1$ and $z=z_2$. Equation \eqref{29102023c} yields
\begin{align}
P^a=-\frac{c^4}{8\pi G}\Delta x\Delta y \left[\left(\frac{\partial e}{\partial z}\right)(z_2)-\left(\frac{\partial e}{\partial z}\right)(z_1)\right]v^a,
\label{15102024a}
\end{align}
which reproduces the result obtained in Ref.~\cite{PhysRevD.108.044043}, Eq. (39) there, for the particular case $f_2=g_2=0$ ($+$ polarization).

In Ref.~\cite{PhysRevD.108.044043}, an analysis of the possible values for the cross-sectional area $A$ suggested that this area has to be discrete in the presence of an electromagnetic wave, for consistency with the hypotheses that electromagnetic waves have discrete energy. Since the cross-sectional area is given by $A=\int\int dx dy \sqrt{^2g}=\int\int dx dy e=e\Delta x\Delta y$, we can recast \eqref{15102024a} in terms of the derivative of $A$ evaluated at the two opposite faces, as in equation (40) of Ref.~\cite{PhysRevD.108.044043}. Hence, the analysis made there holds here as well, and the conclusions about $A$ are the same.

\subsection{Irreducible decomposition of $\rsconnection{_a_b_{(0)}}$}\label{20112024a}
To understand the relation between the teleparallel frame and the energy, it is useful to analyze the properties of the observers' congruence of worldlines. We do that by decomposing $\rcd{_\beta}\e{_{(0)}}$ into its irreducible parts with respect to the rotation group. Given a congruence whose tangent vector field is the $4$-velocity $u=c\e{_{(0)}}$, the components of the Levi-Civita covariant derivative of $u$ is related to $\rsconnection{^a_b_{(0)}}$  by  $\rcd{_\beta} u_\alpha=c\e{_a_\alpha}\e{^b_\beta}\rsconnection{^a_b_{(0)}}$. Using this relation, we can express the decomposition of  $\rcd{_\beta}\e{_{(0)}}$ in the equivalent form \cite{PhysRevD.108.044043}
\begin{align}
\rsconnection{_a_b_{(0)}}=\frac{1}{c}\left(\omega_{ab}+\theta_{ab}\right)+\frac{1}{c^2}a_a\delta^0_b,
\label{28112022e}
\\
\theta_{ab}=\sigma_{ab}+\frac{1}{3}\theta h_{ab},
\label{28112022f}
\end{align}
where\footnote{Notice that $\delta_a^i\delta_b^i=\delta_a^1\delta_b^1+\delta_a^2\delta_b^2+\delta_a^3\delta_b^3$.} $h_{ab}=\delta_a^i\delta_b^i$, $a_a$ is the {\it acceleration vector field}, $\omega_{ab}$ is the {\it vorticity tensor}, and $\theta_{\alpha\beta}$ is the {\it expansion tensor}. The trace-free tensor $\sigma_{ab}$ is the {\it shear tensor}, while the trace of $\theta_{ab}$, denoted by  $\theta$, measures the expansion of the congruence. From $\sigma_{ab}$ and $\omega_{ab}$, one defines $\left[(1/2)\sigma^{ab}\sigma_{ab}\right]^{1/2}$ and $\left[(1/2)\omega^{ab}\omega_{ab}\right]^{1/2}$ as the {\it shear} and the {\it vorticity}, respectively. 

The expansion and vorticity tensors can be calculated from (for details see Sec.~II~D of Ref.~\cite{PhysRevD.108.044043}) 
\begin{align}
\theta_{(i)(j)}=\frac{c}{2}\left(\rsconnection{_{(i)(j)(0)}}+\rsconnection{_{(j)(i)(0)}}\right),
\label{28112022g}
\end{align}

\begin{align}
\omega_{(i)(j)}=\frac{c}{2}\left(\rsconnection{_{(i)(j)(0)}}-\rsconnection{_{(j)(i)(0)}}\right).
\label{28112022h}
\end{align}
(Note that, in the tetrad basis, only the spatial part is nonzero.) From Eq.~\eqref{14082024g} we find that the vorticity vanishes, while the expansion tensor is given by
\begin{align}
||\theta_{(i)(j)}||=-c
	\begin{pmatrix}
	\alpha & \beta\\
	\beta  & \bar{\alpha}
	\end{pmatrix},
\label{10102024a}
\end{align}
where the indices $i,j$ above run over $1,2$ only. Using Eqs.~\eqref{16022024f} and \eqref{14082024e}, we find $\theta=d\ln e/du$, which generalizes equation (30) of Ref.~\cite{PhysRevD.108.044043}. 

Noting that $\theta$ can be recast as $\theta=-(c/e)\partial_z e$, we find that Eq.~\eqref{15102024a} can be rewritten in the form
\begin{align}
P^a=\frac{c^3}{8\pi G} \left[A(z_2)\theta(z_2)-A(z_1)\theta(z_1)\right]v^a,
\label{20112024b}
\end{align}
where $A(z)=e(t-z/c)\Delta x\Delta y$. It is clear that $P^a$ is proportional to the trace of the expansion tensor. As we will see later, the determinant of Eq.~\eqref{10102024a} is proportional to the gravitational energy density.

\subsection{Gravitational and spacetime energy densities}
In order to obtain the gravitational stress-energy tensor, $\energy{^\mu^a}$, the following results are useful:
\begin{align}
\hat{x}_cv^{[c}\hat{x}_{d]}=\hat{y}_cv^{[c}\hat{y}_{d]}=-\frac{1}{2}v_d,
\label{09092024a}
\\
v_d v^{[b}\hat{x}^{d]}=v_d v^{[b}\hat{y}^{d]}=0.
\label{09092024b}
\end{align}
They yield 
\begin{align}
\hat{x}_cv^{[c}\hat{x}_{d]} v^{[b}\hat{x}^{d]}&=\hat{y}_cv^{[c}\hat{y}_{d]} v^{[b}\hat{x}^{d]}=\hat{x}_cv^{[c}\hat{x}_{d]} v^{[b}\hat{y}^{d]}=
\nonumber\\
&=\hat{y}_cv^{[c}\hat{y}_{d]} v^{[b}\hat{y}^{d]}=0.
\label{09092024c}
\end{align}
Given Eqs.~\eqref{14082024g} and \eqref{14082024i}, and Eqs.~\eqref{09092024a}-\eqref{09092024c}, it is straightforward to verify that $\rsconnection{^c_a_d}\rsconnection{^b_c^d}=\rsconnection{^b_a_d}\rsconnection{^c_c^d}=\rsconnection{^b_d_a}\rsconnection{^c_c^d}=\rsconnection{^c_c_f}\rsconnection{^d_d^f}=\rsconnection{^c_d_f}\rsconnection{^d_c^f}=0$, $\rsconnection{^c_d_a}\rsconnection{^b_c^d}=-(\alpha^2+\bar{\alpha}^2+2\beta^2)v_av^b$ and $\rsconnection{^c_c_a}\rsconnection{^d_d^b}=(\alpha+\bar{\alpha})^2v_av^b$. Therefore, Eq.~\eqref{09092024d} gives
\begin{align}
\energy{^b_a}=-\frac{4k}{ec^2}(\dot{f}_1\dot{g}_1-\dot{f}_2\dot{g}_2)v_av^b,
\label{09092024e}
\end{align}
where we have used $\beta^2-\alpha\bar{\alpha}=\beta\bar{\beta}-\alpha\bar{\alpha}=-(\dot{f}_1\dot{g}_1-\dot{f}_2\dot{g}_2)/(ec^2)$. Comparing Eqs.~\eqref{09092024e} and \eqref{10102024a}, we see that $\energy{^\mu^\nu}$ is proportional to the determinant of the expansion tensor.    Equation \eqref{09092024e}  agrees with Eq.~(35) of Ref.~\cite{PhysRevD.108.044043} for $f_2=g_2=0$.

The weak field approximation of Eq.~\eqref{09092024e} agrees with Eq.~(18) in Ref.~\cite{doi:10.1002/andp.201800320} [compare also with Eq.~(1.136) of Ref.~\cite{MicheleMaggiore1}, obtained in a complete different context]. To see this, one just has to take $f_1\approx 1-h_+/2$, $g_1\approx 1+h_+/2$, and $f_2=g_2\approx h_{\times}/2$.

Let us now calculate the spacetime stress-energy tensor. Since we do not know ${\cal T}^{\mu a}$, we use the left-hand side of Eq.~\eqref{29032019k} to calculate $e\tau^{\mu a}$ indirectly, i.e, through the equation $e\tau^{\mu a}=4k\pd{_\alpha}\left( e\potential{^a^\mu^\alpha}\right)$. First, we multiply Eq.~\eqref{05092024a} written in the form $\potential{^a^\mu^\alpha}$ by $e$ and apply $\partial_\alpha$. Then, using Eqs.~\eqref{13082024h} and \eqref{13082024i}, and also the last expression in Eq.~\eqref{14082024b}, we find $\partial_\alpha e\hat{x}^\mu=e(\bar{\alpha}\hat{x}^\mu-\beta\hat{y}^\mu)v_\alpha$ and $\partial_\alpha e\hat{y}^\mu=e(-\beta\hat{x}^\mu+\alpha\hat{y}^\mu)v_\alpha$, which lead to $\partial_\alpha e\hat{x}^\alpha=\partial_\alpha e\hat{y}^\alpha=0$. From these results and the fact that $v^\mu$ is constant, we see that $\partial_\alpha e v^{[\mu}\hat{x}^{\alpha]}=\partial_\alpha e v^{[\mu}\hat{y}^{\alpha]}=0$. In addition, the equality $\partial_\mu u=-(1/c)v_\mu$ and the scalar products \eqref{13082024e} and \eqref{14082024c} imply that $(\partial_\alpha F)v^{[\mu}\hat{x}^{\alpha]}=(\partial_\alpha F)v^{[\mu}\hat{y}^{\alpha]}=0$ for any function $F$ of $u$. Therefore, we are left with $\pd{_\alpha}\left( e\potential{^a^\mu^\alpha}\right)=-\partial_\alpha[e(\alpha+\bar{\alpha})]\hat{t}^{[\mu}\hat{z}^{\alpha]}v^a$. Using Eqs.~\eqref{16022024f} and \eqref{14082024e} we arrive at
\begin{align}
e\tau^{\mu a}=-\frac{c^2}{8\pi G}v^\mu v^a\ddot{e},
\label{26092024a}
\end{align}
where we have used $k=c^4/(16\pi G)$. This result generalizes that of Ref.~\cite{PhysRevD.108.044043}.

\subsection{Matter energy density and field equation}
To take into account all possible polarizations of an electromagnetic wave propagating along the $z$ direction, we can use the general form $A=A_a\teta{^a}$, where $A_a$ are functions of $u$. Nevertheless, the field equations force the component $E_{(3)}$ to vanish, which, in terms of the $4$-potential, is equivalent to $\dot{A}_{(0)}=-\dot{A}_{(3)}$. Choosing $A_{(0)}=A_{(3)}=0$, we find that we only need $A=A_{(1)}\teta{^{(1)}}+A_{(2)}\teta{^{(2)}}$.

From $A=A_\mu dx^\mu$ and $\hat{x}_\mu=\e{^{(1)}_\mu}$, $\hat{y}_\mu=\e{^{(2)}_\mu}$, which hold for the case we are dealing with, we see that the components $A_\mu$ can be written as
\begin{align}
 A_\mu=A_{(1)}\hat{x}_\mu+A_{(2)}\hat{y}_\mu.
\end{align} 
Substitution into Eq.~\eqref{24092024c} yields
\begin{align}
F_{\mu\nu}=2E_{(1)} v_{[\mu}\hat{x}_{\nu]}+2E_{(2)} v_{[\mu}\hat{y}_{\nu]},
\label{25092024a}\\
E_{(1)}= -\dot{A}_{(1)}/c+\alpha A_{(1)}+\beta A_{(2)},
\label{25092024b}\\
E_{(2)}= -\dot{A}_{(2)}/c+\beta A_{(1)}+\bar{\alpha} A_{(2)},
\label{25092024c}
\end{align}
where we have used Eq.~\eqref{24092024d}. [Note that $E_{(1)}$ and $E_{(2)}$ are the components of the electric field in the tetrad basis. This can be verified by using Eq.~\eqref{25092024a} in the form $F_{ab}$ and using Eq.~\eqref{13082024g}.]

To calculate Maxwell energy-momentum tensor, Eq.~\eqref{24092024b}, we use the scalar products given by Eq.~\eqref{14082024c}. From these products, we find $F^{\alpha\beta}F_{\alpha\beta}=0$ and arrive at
\begin{align}
\calt^{\mu\nu}=\epsilon_0\left[E_{(1)}^2+E_{(2)}^2\right]v^\mu v^\nu.
\label{26092024b}
\end{align}

Using $\tau^{\mu a}=\energy{^\mu ^a}+\calt^{\mu a}$ and Eqs.~\eqref{09092024e}, \eqref{26092024a} and \eqref{26092024b}, we can write the field equation in the explicit form
\begin{align}
-\ddot{e}=2(-\dot{f}_1\dot{g}_1+\dot{f}_2\dot{g}_2)+\frac{\epsilon_0 c^2}{2k}e\left[E_{(1)}^2+E_{(2)}^2\right].
\label{26092024c}
\end{align}
In searching for solutions, it may be more convenient to write this equation without the total derivative of $e$:
\begin{align}
-(\ddot{f}_1g_1+f_1\ddot{g}_1)+\ddot{f}_2g_2+f_2\ddot{g}_2=\frac{\epsilon_0 c^2}{2k}e\left[E_{(1)}^2+E_{(2)}^2\right].
\label{27092024a}
\end{align}
Notice that the gravitational energy density is exactly the term we need to add to Eq.~\eqref{27092024a} in order to make the total derivative of $e$ appear. It is also worth noting that if we had used Einstein's field equation in its original form, we would have obtained the left hand side of Eq.~\eqref{27092024a}; it is the TEGR which allows us to interpret the terms in Eq.~\eqref{26092024c} as energy densities.

As an application, let us assume that $E_{(1)}=E_0\cos\omega u$ and $E_{(2)}=E_0\sin\omega u$, where $E_0$ is a constant. In this case, the right-hand side of Eq.~\eqref{27092024a} becomes $\epsilon_0 c^2 e E_0^2/(2k)$. In turn, by assuming that $f_2=g_2=0$ and $f_1=g_1=f$, we obtain $\ddot{f}=-\omega_0^2 f$ with $\omega_0^2=4\pi\epsilon_0 G E_0^2/c^2$. This is basically the same equation as equation\footnote{The $4\pi$ factor in the $\omega_0$ is due to the convention used here for the Maxwell energy-momentum tensor [compare \eqref{24092024b} here with (46) there].} (48) of Ref.~\cite{PhysRevD.108.044043}. However, the result obtained here is more general because  the polarization is circular rather than fixed.

\section{Concluding remarks}\label{04122024c}
In this paper we have shown that, in the freely falling frame considered here, the energy-momentum of the pp-waves is proportional to the trace of the expansion tensor of the observers' worldlines, Eq.~\eqref{20112024b}, while the  determinant is proportional to the gravitational energy density, Eq.~\eqref{09092024e}. These results suggest the interesting possibility that the intrinsic properties of the congruence might carry all the information about both the gravitational and spacetime energies. Of course, the congruence used here was a particular case of timelike geodesics, and the frame was Fermi-Walker transported. So, the simple relations obtained here between the intrinsic properties of the integral curves of $\{\e{_a}\}$ and the energies might not reflect a general property.

We have generalized the results of Ref.~\cite{PhysRevD.108.044043} to the case of $pp$-waves where both $+$ and $\times$ polarizations of the gravitational waves are present, and generalized the Einstein-Maxwell equations to the case in which the polarization of the electromagnetic wave is not necessarily fixed. We were able to give a simple example with a particular circular polarization and find a harmonic oscillator equation. It would be interesting to study more general cases, where, perhaps, the solutions were more realistic and testable.

The frame given by Eqs.~\eqref{13082024b} and \eqref{13082024g}-\eqref{13082024i} satisfies the time gauge, i.e., $\e{_{(i)}^0}=0$. It was proved in Ref.~\cite{PhysRevD.106.044021} that the quantity that is identified as a type of angular momentum vanishes (spatial part) when the teleparallel frame satisfies this gauge. If this identification is right, then that quantity predicts that there is an angular momentum that vanishes regardless of the polarizations of either the electromagnetic or gravitation waves. As argued in Ref.~\cite{Penrose1982}, defining an angular momentum in general relativity is complicated; although it is also complicated in the TEGR, we hope to find a reasonable explanation for the vanishing of the TEGR angular momentum in the time gauge even when we have circularly polarized waves. This issue will be discussed in detail elsewhere.

\appendix*

\section{The relation between our conventions and those of \cite{GriffithsJBWaves}}
The line element \eqref{16022024d} is an adaptation of Eq.~(4.11) of Ref.~\cite{GriffithsJBWaves}, where, in addition to the signature change and the use of $c\neq 1$, we have also changed: $a\to f_1$, $b\to f_2$, $c\to g_1$ and $e\to g_2$. It is also important to emphasize that we use $u\equiv t-z/c$ and $v\equiv t+z/c$.

From the coordinate change $X=f_1x+f_2y$, $Y=g_2x+g_1y$, and
\begin{align}
r=&\ v+R,
\\
R=&\ \frac{1}{c^2}(f_1\dot{f}_1+g_2\dot{g}_2)x^2+\frac{1}{c^2}(f_2\dot{f}_2+g_1\dot{g}_1)y^2
\nonumber\\
&+\frac{1}{c^2}\left[\frac{d}{du}\left(f_1f_2+g_1g_2\right)\right]xy,
\end{align}
where we have changed $r$ to $cr/\sqrt{2}$ and $v$ to $cv/\sqrt{2}$ in Eq.~(4.9) of Ref.~\cite{GriffithsJBWaves}, one can put Eq.~\eqref{16022024d} in the form
\begin{align}
ds^2=&-c^2dudr-\frac{1}{2}\left(h_{11}X^2+2h_{12}XY+h_{22}Y^2\right)du^2
\nonumber\\
&+dX^2+dY^2.
\label{27082024a}
\end{align}
[We have also changed $u$ to $cu/\sqrt{2}$, which has led to the change $d/du \to (\sqrt{2}/c)d/du$.] Our definition of the $h$s is such that they have units of one over time squared, and their relations with $f_1$, $f_2$, $g_1$, and $g_2$ are given by [Eq.~(4.10) of Ref.~\cite{GriffithsJBWaves}]:
\begin{align}
2\ddot{f}_1+h_{11}f_1+h_{12}g_2=0,
\label{26082024a}\\
2\ddot{f}_2+h_{11}f_2+h_{12}g_1=0,
\label{26082024b}\\
2\ddot{g}_1+h_{12}f_2+h_{22}g_1=0,
\label{26082024c}\\
2\ddot{g}_2+h_{12}f_1+h_{22}g_2=0,
\label{26082024d}\\
f_2\dot{f}_1-f_1\dot{f}_2-g_2\dot{g}_1+g_1\dot{g}_2=0. 
\label{26082024e}
\end{align}

It is common to write the line element \eqref{27082024a} in the form\footnote{The exact form may vary from author to author.}
\begin{align}
ds^2=&-c^2\left(1+\frac{H}{2c^2}\right)dT^2+dX^2+dY^2
\nonumber\\
&+\left(1-\frac{H}{2c^2}\right)dZ^2+\frac{H}{c^2}cdTdZ,
\label{27082024b}
\end{align}
where
\begin{align}
H(u,X,Y)=a(u)(X^2-Y^2)+2b(u)XY+c(u)(X^2+Y^2),
\end{align}
and the functions $a(u)$, $b(u)$ and $c(u)$ are given by
\begin{align}
a(u)=\frac{1}{2}[h_{11}(u)-h_{22}(u)],\ b(u)=h_{12}(u),
\label{11102024b}\\
c(u)=\frac{1}{2}[h_{11}(u)+h_{22}(u)].
\label{27082024c}
\end{align}
The relation between the new coordinates  $(T,Z)$ and the old ones $(t,z)$ is $T=t+(1/2)R$ and $Z=z+(c/2)R$. (Note that $u=t-z/c=T-Z/c$ and $r=T+Z/c$.)

In some sense, we can see $a(u)$ and $b(u)$ as being related to the amplitude and polarization of the gravitational wave, while $c(u)$ is related to the amplitude of the electromagnetic wave, as will become clear below.

Since the amplitudes of the waves can be written in terms of the $h$s, it is interesting to solve Eqs.~\eqref{26082024a}-\eqref{26082024e} for them. This gives
\begin{align}
h_{11}=-2(g_1\ddot{f}_1-g_2\ddot{f}_2)/e,\ h_{12}=2(f_2\ddot{f}_1-f_1\ddot{f}_2)/e
\nonumber\\
h_{22}=-2(f_1\ddot{g}_1-f_2\ddot{g}_2)/e.
\label{11102024a}
\end{align}
By substituting Eq.~\eqref{11102024a} into Eqs.~\eqref{11102024b} and \eqref{27082024c}, we obtain
\begin{align}
a(u)=\frac{1}{e}\frac{d}{du}\left(f_1\dot{g}_1-\dot{f}_1g_1+\dot{f}_2g_2-f_2\dot{g}_2\right),
\label{11102024c}\\
b(u)=\frac{2}{e}\frac{d}{du}\left(\dot{f}_1f_2-f_1\dot{f}_2\right),
\label{11102024d}\\
c(u)=-\frac{1}{e}\frac{d^2e}{du^2}+\frac{2}{e}\left(\dot{f}_1\dot{g}_1-\dot{f}_2\dot{g}_2\right).
\label{11102024e}
\end{align}

Comparing Eqs.~\eqref{26092024c} and \eqref{11102024e}, we find that $c(u)=(\epsilon_0c^2/2k)\left[E_{(1)}^2+E_{(2)}^2\right]$.

Equation \eqref{11102024d} shows that if $f_2$ (or $g_2$) vanishes, then $b=0$ (no cross polarization for a gravitational wave).

\section*{Acknowledgments}
Costa, R. D. acknowledges CAPES for financial support.

%


\end{document}